# Does Quantum Cosmology Predict the Age of the Universe?

**Álvaro Mozota Frauca**[1]

**Abstract**
The problem of time of quantum gravity has been argued to make canonical approaches unsatisfactory. In this article I study how it affects quantum cosmology and reach the same conclusion. The advantage of studying the cosmological case is that its simplicity makes the discussion much clearer and less technical. The classical models I will be concerned with describe how two degrees of freedom, the scale factor and a scalar field, evolve with respect to a time variable. After quantizing the model, this time variable just disappears, and I argue that this is problematic. Indeed, this variable in the classical model allowed us to make claims like 'the universe is 13.8 billion years old' and I will argue that these claims are physically meaningful predictions that are lost in quantum cosmology. I will analyze some of the relational positions in the quantum gravity and quantum cosmology literature that tend to deny the physical meaning of time variables and I will argue against them for the case of classical cosmology. I conclude that the age of the universe is a physical prediction of classical cosmological models, that it is missing from quantum cosmology, and that this should make us suspect that there is something wrong with this sort of approach.

## 1 Introduction

Quantum cosmology builds cosmological models by taking into account quantum effects that could alter our picture of the Big Bang and the beginning of the universe. In this area of study many philosophically interesting questions arise about the nature of space and time, about how to do quantum mechanics for a whole universe, and about the origin and fate of our universe. In this paper I will focus on an objection that one can raise against the most

✉ Álvaro Mozota Frauca
  alvaro.mozota@upc.edu

1  Universitat Politecnica de Catalunya, Av. Diagonal, 649, 08028 Barcelona, Spain

relevant class of quantum cosmological models, those which are built by canonical quantization techniques. This objection is that, as I will argue, these models leave out an important part of the empirical content of classical cosmological models, namely, everything related to time, duration, and temporality. For instance, the age of the universe is a prediction[1] of classical cosmological models that goes missing.

The reason for this is that canonical quantization famously leads to what is called the problem of time when applied to theories like general relativity. The reparametrization invariance of some classical theories, including cosmological models, makes it the case that quantization fails to give a non-trivial dynamical equation. The dominant view in the quantum gravity and cosmology literature is that this problem can be circumvented and that there are some ways of taking the outcome of such a quantization procedure to constitute a physically sensible model. However, here I build on previous criticisms against this view, I will argue that the problem of time makes canonical quantum cosmological models fundamentally flawed, and I will illustrate this with the fact that they are unable to accommodate predictions about time, such as the age of the universe. For doing so, I will have to argue against some objections that can be raised from relationalist points of view that tend to deny the physical meaning of time.

In this sense, the argument in this paper is related to the more complex discussion concerning the problem of time for canonical approaches to quantum gravity. The advantage of taking quantum cosmology as a case study is its simplicity, which will allow for a clearer discussion without entering into the sometimes technical details of general relativity, its gauge and temporal structures, and its quantization. In particular, while discussing the physical and empirical content of general relativity is nuanced and requires some technical knowledge, it will hopefully be less controversial and more easily understandable to claim that cosmological models make empirical claims about the age of the universe. By arguing that this sort of empirical content is missing in quantum cosmology, I am offering a complementary argument to those arguing that the problem of time seriously affects canonical quantizations of general relativity.[2]

Let me clarify that my argument is not that every concept or structure of classical cosmology should be present also in quantum cosmology. Every time there is a theory change, there is also a change in theoretical structures, such that some of the concepts of the old theory make limited sense in the new theory. In this sense, it is possible that in a theory of quantum gravity or quantum cosmology, our concepts of spacetime need to be updated or even dropped. However, even if this is the case, the new theory should have the resources to account for the empirical content of the old theory. For instance, when replacing Newtonian mechanics with a relativistic theory one needs to drop the idea of absolute simultaneity and instantaneous force, but the theory is still able to account for interactions by introducing the appropriate fields to mediate in the interactions. The empirical content of the old theory is therefore preserved (maybe with small amendments).

---

[1] I am thankful to an anonymous reviewer for pointing out that the term 'prediction' may not be entirely adequate for the case of the age of the universe. The point I want to make when I use the term 'predict' is that it is part of the empirical content of cosmological models and that the fact that the universe has a finite age of 13.8 billion years can be seen as an expected consequence of our physical models. Still, if the reader thinks that terms like 'retrodict' or 'explain' are more adequate, they can read my claims as referring to retrodictions or explanations.

[2] In particular, this article can be seen as complementary to Mozota Frauca (2023), where the case of general relativity is discussed paying more attention to technical detail.

In the case of the canonical quantum cosmological models that I will present in this article we have a formalism in which temporal structure is missing. This wouldn't be problematic if we had an interpretation for this formalism and its models which would be able to explain the empirical content of the classical model that we relate to time. However, I will argue that in the main two interpretations that we have of quantum cosmological models, the internal clock interpretation and the probabilistic interpretation, it seems that we are not able to recover this empirical content, not even by invoking some new understanding of space and time or a lack of them.

The structure of this paper is the following. First I start in Sect. 2 by briefly introducing simple classical FLRW cosmological models, the way they have been proposed to be canonically quantized, and the way time appears to be missing. In Sect. 3, I give a summarized version of the argument in Mozota Frauca (2023; 2024) that argues that the problem of time makes the canonical quantization of non-deparametrizable reparametrization invariant models unsatisfactory[3] and that this affects general relativity and I will argue that it also affects the quantization of classical cosmological models. Then, in Sect. 4, I argue, against some relational views popular in the literature, that the predictions about duration, such as the age of the universe, are physically meaningful predictions of classical cosmological models and I argue that it is problematic that they are missing from the quantum model. Finally, in Sect. 5, I conclude.

## 2 Cosmology and Quantum Cosmology

### 2.1 Classical Cosmology: Minisuperspace Models

Contemporary cosmological models have reached an impressive level of complexity and completeness for the phenomena they describe. They are able to accommodate all kinds of matter, dark matter, dark energy, inhomogeneities and anisotropies, quantum fluctuations, and different inflationary mechanisms, to mention a few. Here I will give just a brief introduction to the simplest model, which is an FLRW spacetime filled with a scalar field. This model contains the basic features of cosmological models and is customarily studied in introductory-level cosmology textbooks.[4] Moreover, this type of model is the one that has been most commonly used as a starting point for building quantum cosmological models. These models are known as minisuperspace models, as they are defined on a configuration space which is a reduced version of superspace, the configuration space of general relativity.

Contemporary cosmology is a relativistic theory, and, therefore, one has to be cautious when discussing issues regarding time in this setting. In relativity it is well-known that each physical system experiences a different time, and that different clocks traveling through spacetime measure different times depending on their trajectory in spacetime. However, as the universe we live in is, to a very good degree of approximation, quite symmetric, we are able to find a well-defined and convenient way of dividing spacetime into space and time, that is, to introduce a foliation of spacetime. Relative to this foliation, galaxies are approximately at rest, which means that the clocks at these galaxies run at the same pace.

---

[3] Notice that it is just the symmetry of the models which is responsible for this issue. That is, even for models which have nothing to do with gravity or dynamical spacetimes we find this issue.

[4] See for instance Weinberg (2008, Ch. 1.12).

The time that a clock in any of these galaxies measures approximately is called cosmic time. In a model with an initial Big Bang singularity, one can define the age of the universe to be the cosmic time that has elapsed since the initial singularity until our days. The age of the universe is an empirically meaningful part of our cosmological models: our universe is believed to be 13.8 billion years old, and if it were younger or older, what we would observe in the galaxies around us would be different. In this sense, even if cosmology is a relativistic theory, for practical purposes it does make sense to introduce a preferred, conventional way of splitting spacetime into space and time and to speak about the age of the universe.[5]

The idea of cosmic time and the age of the universe is even more powerful and convenient when we study the very early universe. According to our best theories and evidence, right after the singularity, the universe was a hot soup of particles that cooled down as it expanded. In this very early universe, we can also introduce a foliation of spacetime that is convenient and that makes everything simple. Relative to this foliation, the properties of this soup of particles (composition, density, temperature) are homogeneous to an astonishing degree. Using this fact, our cosmological models are able to describe how these properties evolve with respect to cosmic time in a simple and powerful way. It is in this sense that cosmologists make claims about, say, the time that took the universe to cool down enough for atoms to form. Again, facts about duration are an important part of the empirical content of these models. For instance, the abundance of helium in our universe depends on how long given epochs of the early universe last. In particular, this abundance depends on the quantity of neutrons at the time of nucleosynthesis, and these neutrons are the ones that have 'survived' the epoch in which no more neutrons are created while some of them decay. If this epoch had been longer, fewer neutrons would have reached the time of nucleosynthesis, and today we would observe less helium. Conversely, if it had been shorter, the abundance of helium would be higher today.

Another important conceptual issue with general relativistic theories has to do with the difference between coordinate time and proper time. In general relativity, the coordinates one uses to identify points in spacetime can be chosen freely, which means that they are not straightforwardly associated with what clocks and rods will measure. However, since in cosmology we are able to introduce the convenient split between space and time, we can work directly with physically meaningful coordinates. That is, instead of introducing a general time coordinate $\tau$, we can work with models in which the time coordinate is $t$, the cosmic time that clocks at typical galaxies measure. For this reason, cosmology is a case in which even if the theory is general relativistic, the physical meaning of the coordinates employed is straightforward. However, at a formal level cosmology still allows us to choose freely the set of coordinates we want to use, and this will be what will cause the problem of time at the time of quantizing cosmological models. That is, as the model is written in terms of an arbitrary parameter $\tau$ and not in terms of $t$, it will have the kind of symmetry that is troublesome at the time of quantization.

After this conceptual discussion, we are ready to discuss minisuperspace models. As I have just mentioned, our universe is very symmetric, in particular, when studying it at cosmological scales it looks very flat, homogeneous, and isotropic, especially for the very early universe. For this reason, it is a good approximation to use a spacetime metric describing a universe with these properties. This spacetime is a flat Friedmann-Lemetre-Robertson-Walker (FLRW) spacetime, which can be described by the following line element:

$$ds^2 = -dt^2 + a^2(t)(dx^2 + dy^2 + dz^2).\tag{1}$$

---

[5] For more extensive discussions of how to think about time in cosmology and about its philosophical implications I refer the reader to Smeenk (2013); Callender and McCoy (2021).

As I have mentioned above, this metric introduces a natural foliation of spacetime: there is a way of dividing into space and time such that at every moment of time $t$ space is described as being homogeneous. The time coordinate $t$ is cosmic time: it is the proper time an observer at rest with respect to this foliation would measure. The only non-fixed function in the FLRW metric is the scale factor, $a(t)$, which intuitively describes how 'big' the universe is. A growing $a(t)$ represents an expanding universe, a decreasing $a(t)$ a contracting one, and if it is 0 we have a singularity like the Big Bang or Big Crunch singularities that have been studied in the cosmology literature.

The dynamics of the geometry of the universe is reduced to the dynamics of $a(t)$ for such a simple model, and, in particular, the 10 Einstein equations of general relativity are reduced to just two independent equations, the Friedmann equations. These equations relate the evolution of the scale factor of the universe with its matter content, just as the full Einstein equations related the curvature of spacetime with the stress-energy tensor of matter. Given the symmetries of the model, in this cosmological setting the stress-energy of matter is described by just two functions, the energy density $\rho$ and pressure $P$. Realistic models include different types of matter, as well as dark matter and dark energy that contribute to $\rho$ and $P$, and they arrive at a good level of prediction.

For simplicity, and because it is the way basic quantum cosmological models are built, we will just take the matter content of the universe to be described by a single scalar field $\phi$. By specifying its dynamics in some of the ways it is done in the literature[6] we would have completed the construction of our classical cosmological model. Solutions of the equations of motion specify pairs $a(t), \phi(t)$, which are enough for representing some important cosmological information. For instance, such a model can predict an expanding universe which is to an extent in agreement with cosmological observations. And, most importantly for the argument in this paper, it predicts an initial singularity that happened some time $t_0$ ago, which would be the age of our universe. Of course, the time $t_0$ in this simple model corresponds to just an approximation to the real age of our universe, that our best cosmological models and observations estimate at around 13.8 billion years.[7]

An important feature of many of these models for the later discussion is that both $a$ and $\phi$ are monotonic in time. This allows using them for identifying instants of the evolution, i.e., to use them as clocks. In this sense, we can express part of the dynamics of the model as the correlations $a(\phi)$ or $\phi(a)$. This allows us to make and answer questions like what the scale factor of the universe is when the scalar field takes some given value. However, even if it also allows us to formulate questions about how the scalar field changes in the time that takes the universe to double its size, it doesn't capture how long that time is. In this sense, studying $a(\phi)$ or $\phi(a)$ may be interesting, but it doesn't represent all the physically meaningful predictions of the model. In particular, the prediction of the age of the universe is not contained in $a(\phi)$ or $\phi(a)$.

---

[6] See for instance the discussions of scalar field models in Weinberg (2008); Ashtekar and Singh (2011); Calcagni (2017).

[7] See for instance the results of the recent Planck collaboration (Ade et al. 2016; Aghanim et al. 2020), which build complex and complete models based on precise cosmological observations. These results estimate the age of the universe to be 13.8 billion years with an accuracy of the order of 0.03 billion years (Ade et al. 2016, Table 4).

This ends this brief discussion of the classical model. To discuss its quantization and the problem of time associated with it, we need to explicitly introduce its canonical formulation using the Hamiltonian formalism. It is customary to preserve part of the full diffeomorphism invariance of general relativity, which is the part that corresponds to temporal reparametrization. It is because of this reparametrization invariance that one needs to introduce the constrained formalism to deal with this symmetry of the theory and that the problem of time arises.[8] The (total) Hamiltonian of the theory is of the form:

$$H(a, p_a, \phi, \pi, N, \pi_0) = N\mathcal{H} + \lambda(\tau)\pi_0, \tag{2}$$

where $\mathcal{H}$ is a phase space function known as the Hamiltonian constraint, $N$ is the lapse function, $p_a, \pi$, and $\pi_0$ the momenta conjugate to $a, \phi$, and $N$, and $\lambda(\tau)$ an arbitrary function. This Hamiltonian specifies how all phase space functions evolve with respect to an arbitrary time parameter $\tau$,[9] which is related to the physical[10] time parameter by means of $dt = N(\tau)d\tau$. Different functions $\lambda(\tau)$ correspond to different but equivalent time parametrizations, as the solutions of the equations of motion are reparametrization-equivalent and represent the same physics, that can be described by $a(t), \phi(t)$. Moreover, consistency requires that $\mathcal{H}$ and $\pi_0$ vanish during the whole evolution, and for this reason, they are referred to as constraints. The dynamics defined by the Hamilton equations and the constraints are equivalent to the dynamics as defined by Friedmann equations and the equations of motion for the scalar field.

## 2.2 Canonical Quantum Cosmology

Models of canonical quantum cosmology are built by applying canonical quantization methods to classical models like the one just discussed. As the classical theory is a constrained system, one needs to apply Dirac's quantization procedure or some equivalent or similar quantization.[11] I will follow Mozota Frauca (2023) in describing the steps of this quantization process in the following way:

1. Start with the classical theory defined on a constrained phase space.

---

[8] See Rothe and Rothe (2010) for a textbook on constrained systems and how to deal with them in the Hamiltonian formalism, and Mozota Frauca (2023) for a discussion specific to reparametrization invariant systems and general relativity.

[9] In terms of this coordinate the line element (Eq. 1) can be written as $ds^2 = -N^2(\tau)d\tau^2 + a^2(\tau)(dx^2 + dy^2 + dz^2)$. This corresponds to an ADM decomposition of spacetime in which the shift vector is forced to be zero so that homogeneity and isotropy are preserved.

[10] As commented above, it is physical in the sense that it is the time that observers at rest with respect to homogeneous space would measure. Below I will expand on the physicality of $t$.

[11] For the main argument in this article it won't be relevant whether the quantum model is derived by following strictly Dirac's methods as originally formulated in Dirac (1964) or some alternative method like a path integral method as long as it produces a similar outcome in which $t$ disappears.

2. Choose a subalgebra of functions on phase space and quantize them, i.e., build an algebra of operators on a kinematical Hilbert space $\mathcal{H}_{kin}$ such that their commutator algebra is defined by the Poisson algebra of the classical functions.

3. Impose the constraints. That is, define the physical Hilbert space $\mathcal{H}_{phys}$ as the space of the states which satisfy $\hat{C}_A|\psi\rangle = 0$.

4. Build a Hamiltonian operator which is a quantization of the total Hamiltonian that generates the constrained dynamics in the classical theory. The dynamics of the theory is contained in the Schrödinger equation for that Hamiltonian or in some equivalent form.

We have already completed step number one in the previous subsection: we have defined a classical model on a constrained phase space and we have specified its Hamiltonian. Different models of quantum cosmology differ slightly on the variables used or the exact form of the Hamiltonian in the classical model they base their quantization on. The two most relevant types of variables used are metric variables like the scale factor introduced in the previous subsection and connection variables. This latter set of variables is defined in analogy with the connection variables used for the loop quantization of full general relativity in loop quantum gravity (LQG), and the quantization it defines is known as loop quantum cosmology (LQC).[12] Choosing metric or connection variables is equivalent at a classical level and although it has some consequences at the quantum level, for the argument in this paper it won't be relevant, as approaches built choosing both kinds of variables are equally affected by the problem of time.

Step number two is to define a Hilbert space in which a subalgebra of the original phase space is represented by means of operators. Different quantum cosmological models differ in the way this space is defined. For models based on metric variables the definition of the Hilbert space is completely analogous to the definition of the Hilbert space of a single particle in standard quantum mechanics, and this quantization is sometimes referred to as the Wheeler-deWitt quantization, as it is analogous to the Wheeler-deWitt quantization of full general relativity.[13] For models based on connection variables, the definition of the Hilbert space follows a route analogous to the quantization of LQG, and one defines a Hilbert space in which there is a fundamental discreteness. At the end of the day, this implies that in LQC the scale factor of the universe is allowed to take only values from a discrete set.[14] This has sometimes been argued to imply that time is discrete, although a consequence of my argument in my paper will be that this does not follow because of the problem of time. The topic of discreteness and its motivation are philosophically interesting, but they won't play a role in the discussion in this article.

---

[12] I refer the reader to Rovelli (2004) for an introduction to LQG and a discussion of connection variables in general and to Ashtekar (2009), Ashtekar and Singh (2011), Agullo and Singh (2017) for discussions of LQC and the variables used in these models.

[13] This nomenclature is common in the LQC literature, in which they want to establish a clear distinction between quantizations based on metric and connection variables. See for instance (Bojowald and Morales-Técotl 2004; Ashtekar et al. 2006a; 2006b; Bojowald 2011; 2015).

[14] From a technical point of view things are a little bit subtler than my basic exposition here. The basic Hilbert space of LQC cosmology allows for any value of the scale factor, but given the Hamiltonian constraint of the model, it can be decomposed into sectors that are dynamically isolated from each other. For this reason, dynamics can be defined in just one sector, which means that the scale factor takes values from just one discrete set. I refer the reader to the LQC literature for discussions of this point (Ashtekar et al. 2006b; Ashtekar and Singh 2011; Agullo and Singh 2017).

Step three defines the physical Hilbert space of the system, that is the space of states that satisfy the constraints. For instance, in the case of electromagnetism, the kinematical Hilbert space allows for any configuration of the electromagnetic field, while the constraints that define the physical Hilbert space restrict physical states to be those that satisfy Gauss law. In the case of cosmology we had two constraints. The quantum version of the constraint $\pi_0$ implies that physical states are independent of the lapse function $N$, while, as the classical Hamiltonian constraint $\mathcal{H}$ implied the first Friedmann equation, its quantum form is the quantum counterpart of this equation.[15] This defines physical states $\psi(a, \phi)$.

It is in step 4 that we find the problem of time. The quantization of the Hamiltonian 2 would lead to the following Schrödinger equation:

$$\partial_\tau \psi(a, \phi, N, \tau) = (\hat{N}\hat{\mathcal{H}} + \lambda(\tau)\hat{\pi}_0)\psi(a, \phi, N, \tau).\qquad(3)$$

But when we restrict the application of this equation to states that satisfy the constraints, that is, $\hat{\mathcal{H}}\psi = 0$ and $\hat{\pi}_0\psi = 0$, we find that this equation is reduced to just a trivial equation:

$$\partial_\tau \psi(a, \phi, \tau) = 0 \implies \psi = \psi(a, \phi).\qquad(4)$$

That is, contrary to our expectations, we get states that are independent of any time variable. This means that physical states lack any temporal dependence, and I will argue that this is problematic and that a symptom of this is that these states do not contain information about an important part of the classical model, which is everything related to time and temporal durations. This of course includes information about the age of the universe.

Mimicking the mainstream attitude in the quantum gravity community,[16] the fact that the expected-to-be dynamical equation of the model is trivial is considered not to be problematic and the Hamiltonian constraint equation is reinterpreted as a dynamical equation by the quantum cosmology community. In this sense, states in the physical Hilbert space are not considered to be states to be evolved in time but states that already describe how things develop. By studying states in this space, or operators, the community believes that one has a complete quantum model and that predictions can be drawn from it.

The most extended interpretation of states in the physical Hilbert space of canonical approaches to quantum cosmology is the internal clock interpretation,[17] which considers one of the variables $a$, $\phi$ to act as an internal time. If one takes $a$ to be the internal time, $\psi(a, \phi)$ would be describing a wavefunction in which the probability of finding a value for the scalar field is changing while the universe goes from smaller to bigger sizes as reflected by $a$. Alternatively, if one takes $\phi$ to be the internal time variable, the state $\psi(a, \phi)$ represents how the probability distribution of finding different sizes of the universe change for different moments of $\phi$-time, which are identified just by a value of $\phi$. In this article I will be criticizing this view, as I will argue against the claim that $\phi$ or $a$ represent an internal time. In a classical minisuperspace model, $a$ and $\phi$ are dynamical variables that evolve with respect to the physical time parameter $t$, and I will take the fact that in the quantum model this parameter disappears as a sign that there is something worrisome about the quantization of the system.

---

[15] This is in analogy with the canonical quantization of general relativity, where the quantum version of the Hamiltonian constraint defines physical states and is known as the Wheeler-deWitt equation.

[16] See for instance, Kiefer (2012), Rovelli (2004), Rovelli and Vidotto (2022).

[17] This seems to be true for both early approaches (Halliwell 1990) and more recent ones (Ashtekar and Singh 2011; Bojowald 2015; Agullo and Singh 2017; Gielen and Menéndez-Pidal 2022a).

In addition to the main worry in this article, let me also note that the internal time view suffers from the problem that it isn't very clear which of the variables in the formalism we should take to represent internal time. Earlier works in quantum cosmology[18] took the scale factor $a$ to represent the internal time, and there are some reasons for holding this belief. A couple of them are: first, that taking the scale factor to be the internal time is a choice that is independent of the details of the matter fields in play, and second, that the way the scale factor appears in the action of superspace models is similar to the way some time variables appear in parametrized models. However, none of these arguments is definitive and the quantum gravity community, especially the LQC community, shifted to take matter fields like $\phi$ to represent an internal time.[19] There are also some reasons for this, some more technical regarding the way the constraints are imposed or the definition of an appropriate sense of unitary dynamics, and others more conceptual, like the fact that one of the goals of quantum cosmology would be to study whether the prediction of a singularity in the sense of a vanishing scale factor at the beginning of the universe is still valid. For this, one needs to take $a$ as a dynamical variable and not as a time parameter.

Choosing different variables to be the internal time leads to different physical theories and interpretations. Let me cite the recent discussion in Gielen and Menéndez-Pidal (2022a) as showing this, although this problem has been known since the early days[20] of quantum cosmology and quantum gravity. In Gielen and Menéndez-Pidal (2022a; 2022b), the authors work with a classical model slightly more complicated than the minisuperspace model presented here, as they build a model with yet another degree of freedom. From a classical perspective, Gielen and Menéndez-Pidal (2022a; 2022b) take the point of view that any of the three variables in the theory represents an equally valid clock or internal time, and they study the three quantum theories that arise by interpreting each of these clocks as a time variable. For each of these theories, they consider some nice, peaked states and study their behavior. The results are as disparate as they could get: one of the options describes a contracting and then expanding universe (avoiding any singularity), another describes an expanding and then contracting one (reaching a maximum volume), and the third one describes a universe that behaves very similarly to the classical one and which does not avoid the Big Bang singularity.

The position I am taking in this article is that the multiple-choice problem reflects the fact that dynamical variables, even if monotonic, are different from a conceptual point of view from a time parameter and that something problematic happens when applying canonical quantization methods to minisuperspace models. In this sense, I won't be evaluating the arguments for preferring one clock variable over another or one theory over another, as I will consider that the identification of any of them with time is equally wrong from a conceptual point of view, as I will further argue below. Similarly, I agree with the conclusions in Gielen and Menéndez-Pidal (2022a) that the different theories for different choices of clock are indeed different and not equivalent in some sense. In any case, let me mention that even if the reader doesn't agree with my analysis and thinks that there is one privileged internal clock, or that all choices are somehow equivalent, the main argument in this article will still be valid as I will argue that there seems to be no relation between $t$, which I argue has physical meaning, and the internal clock[21].

---

[18] See the review (Halliwell 1990) that covers works from the 1960s till the 1990s.

[19] Early works in LQC were still based on taking geometry to be the time variable (Bojowald and Morales-Técotl 2004), but they shifted to take matter field as time variables (Ashtekar et al. 2006a; 2006b), position that is still held (Agullo and Singh 2017).

[20] See for instance (Unruh and Wald 1989; Kuchar 1992).

Besides the internal clock interpretative scheme, a good deal of the quantum cosmology literature has dealt with semiclassical states and approximations.[22] There is a family of states, WKB states,[23] which are approximate solutions of the constraint equation, at least for some regions of minisuperspace, i.e., the $a - \phi$ plane, and they are very peaked along classical trajectories. Given this property, they have been used for claiming that the original correlation $a(\phi)$ or $\phi(a)$ is contained in the quantum theory in the appropriate limit. Moreover, using these states it has been claimed that one can recover time. A rough way of doing so is directly by postulating that the time parameter associated with the trajectory $a(\phi)$ in minisuperspace needs to be the same time parameter that one would associate to in the classical theory. In this way, one just postulates that in the quantum theory, if one can define something like a trajectory in configuration space, then the time associated with that trajectory needs to be the same that one would assign to the same trajectory in the classical theory. In practice, this is done in a more sophisticated way and one defines 'WKB time' by identifying something analogous to a Hamilton principal function in the phase of states of the WKB form. This allows assigning a time parameter to the trajectory in configuration space associated with that WKB state in a way analogous to the way one would ascribe a time parameter to a classical trajectory knowing the Hamilton principal function of the trajectory. For any semiclassical trajectory, WKB time acts like an internal time and one can define it as a configuration space function. This allows for studying interesting semiclassical effects when more degrees of freedom are added to the model. Moreover, the trajectory in configuration space does not need to be exactly the trajectory of the classical model and the Hamilton principal function derived from the phase of the state does not need to be exactly the same as the one of the classical model. This means that WKB semiclassical cosmology can show interesting departures from the classical models one started with.

---

[21] In models based on unimodular gravity, such as the ones considered in Gielen and Menéndez-Pidal (2022a), one can argue that there is a preferred candidate for internal clock, which is the variable conjugate to the cosmological constant. Unimodular gravity is an extension of general relativity built in order to solve its problem of time (Brown and York 1989). In this sense, adopting a cosmological model based on unimodular gravity for solving the problem of time is acknowledging that standard minisuperspace models as the one I am discussing in this article suffer from this problem and that they need an expansion in order to solve it. It is beyond the scope of this paper to analyze the extent to which unimodular gravity succeeds or not in solving the problem, and I will refer the interested reader to the criticisms of this approach that were raised in Kuchař (1992); Isham (1993). In connection with the discussion here, let me just point out that the time parameter in unimodular cosmology is related to the volume of the universe and not to proper time. In this sense, the recovery of the age of the universe would be still non-trivial in this kind of approach. I will come back to this point in the conclusion (Sect. 5).

[22] See Chua and Callender (2021) for a complementary and also critical view about semiclassical 'resolutions' of the problem of time. See also Huggett and Thebault (2023) for a more sympathetic view of semiclassical cosmology, but one which shares a key point with my analysis here, namely that the temporal structure is just postulated in these approaches.

[23] See the discussions in Unruh and Wald (1989); Kiefer (2012) for an introduction to these states and their applicability in quantum cosmology. The discussion here is based on the more recent literature, although it also applies to early semiclassical approaches as the one proposed in Gerlach (1969).

Does this mean that WKB states constitute counterexamples to my claim that there is something worrisome with canonical quantum cosmology and that a sign of this is that it doesn't predict the age of the universe? I believe they do not for a couple of reasons.

First, my claim concerns full quantum cosmology and not semiclassical approximations or effective theories. In these approaches one is not finding some time that is hidden somewhere in the formalism of quantum cosmology, but one is just postulating this temporal structure.[24] If one accepts that one needs to postulate temporal structure, then it seems that one is accepting my claim that quantum cosmology lacks it. One could also worry about how justified this postulate is, and there is certainly more to say about how sound the interpretation of semiclassical states is, but this discussion is beyond the scope of this article. Let me just say that from my perspective, as I find that canonical quantum cosmological models are problematic and may fail to constitute meaningful models, I am quite skeptical that it is sound to build semiclassical theories based on such foundations.

Second, the WKB time construction is only valid for states that are approximately of the WKB form. When we move to consider generic solutions of the constraint equation, all the WKB machinery ceases to be applicable: we no longer have states peaked along a one-dimensional region of configuration space and we no longer can read an approximate Hamilton principal function from the state. In this sense, even if we thought that adding some postulate like the one in semiclassical cosmology would help us find (or impose) temporal structure in our models, the semiclassical way of doing so is not available. For this reason, semiclassical strategies cannot work to solve the problem of time of the model.

In connection with WKB states, let me mention that an important part of the quantum cosmology literature has dealt with approximate or effective theories. For instance, in the LQC literature (Agullo and Singh 2017; Bojowald 2015) it is common to see studies of a classical minisuperspace model with a corrected Hamiltonian constraint. This constraint has additional terms that are introduced motivated by the belief that space is discrete, and has the effect that the classical dynamics it defines describes a bouncing universe, i.e., a universe that was originally contracting until it reached a minimal size and then started expanding again, which would be the phase of the universe at which we are right now. Models of this sort clearly have a time parameter that allows making claims about time and duration. But in the same way that I claimed that WKB states do not constitute counterexamples to my main claim, these models do not contradict this claim, as my claim concerns canonical quantum cosmology and not effective models. That is, my claim is that even if effective models may have time and make predictions about it, canonical quantum cosmology, as it is defined, does not.

A different kind of interpretation of quantum cosmological states reads them as encoding some probabilities. I won't enter into the detail of such interpretations but it is clear that if we claim that $\psi(a, \phi)$ encodes some sort of probability density of finding the universe with a size $a$ and scalar field $\phi$ we are not making any claim about the time $t$, about the duration of intervals or about anything along those lines. In this sense, in the probabilistic interpretation one can also argue that time has gone missing. I refer the reader to Unruh and Wald (1989); Kuchar (1992); Isham (1993) for critical introductions to this sort of interpretation and to Rovelli and Vidotto (2015; 2022) for a more recent and slightly different interpretation. I'll come back to probabilistic interpretations and how they are affected by the problem of time in Sect. 4.

---

[24]Again, see the discussion in Huggett and Thebault (2023), where this is acknowledged.

Let me wrap up this section by insisting on its main ideas. Canonical quantum cosmological models are based on physical states, states that satisfy a certain Hamiltonian constraint in the space of wavefunctions defined on minisuperspace, that is, in wavefunctions of the form $\psi(a, \phi)$ for the model discussed above. The problem of time of canonical quantizations of reparametrization invariant systems shows up in that these states are independent of the time variable $t$ which was present in the classical model and which had a physical meaning. I have mentioned the main ways in which the quantum cosmology community has interpreted these states, and I have noted that they either leave the predictions concerning $t$ aside or that they are built in some semiclassical approximations that are not valid for generic quantum states. In the following section I will connect the case of quantum cosmology with the discussion of the problem of time for full general relativity and other reparametrization invariant systems.

## 3 Analyzing the Problem of Time

The problem of time arises in the canonical quantization of every (at least temporal) reparametrization invariant model. In this section I will give an overview of the problem as presented in Mozota Frauca (2023; 2024),[25] and I will agree with that analysis in that I will consider that the distinction between deparametrizable and non-deparametrizable models is crucial, and that while for the former class the problem can, in principle, be solved, for the latter the problem is much more severe.[26] Minisuperspace models, just like general relativity, belong to this latter class, and as such the problem of time seems to make the quantization unsuccessful. In particular, I am highlighting that the information about duration, including the age of the universe, is lost in the canonical quantization of these models.

Let me start by introducing the definition of deparametrizable models. A classical, reparametrization invariant model is deparametrizable if one of the variables in its configuration space is a time variable, or if $n$ field variables correspond to $n$ spacetime coordinates for the case of field theories. As in this configuration space one can distinguish between the 'true' configuration variables and the time variable or the spacetime coordinates, one says that this configuration space is an extended configuration space. That is, it is a configuration space that can be built by expanding the 'true' configuration space by the addition of a time variable or of spacetime coordinates.[27]

The classical example of this class of models is the parametrized version of the dynamics of a Newtonian particle. In this example, one describes the evolution of the position of a Newtonian particle with respect to an arbitrary temporal parameter $\tau$. To recover the dynamics, Newtonian time is included in the configuration space of the model, which now is described by a pair $x$, $t$. By studying a trajectory in the extended configuration space $x(\tau), t(\tau)$ one can recover the same information of the original theory, that is, the evolution of the position of the particle with respect to the absolute Newtonian time, $x(t)$. This process is known as deparametrization, as one is eliminating the arbitrary parameter $\tau$ to express physics with respect to $t$, the true time coordinate of the theory.

---

[25] I also recommend the reader the classical reviews of the problem of time (Kuchar 1992; Isham 1993) and the more recent book (Anderson 2017). According to the analysis in that book, the aspect of the problem of time I am focusing on most in this article is the 'frozen formalism' problem.

[26] Notice that even if the problem of time is generally discussed in the context of gravitational theories, the reason why it arises is the symmetry of these theories. For this reason, this problem also arises in the quantization of models that have nothing to do with gravity, such as the example I discuss in this section.

[27] Notice that there is some disagreement in the way I use the term 'extended configuration space' and the way it is employed by authors like Rovelli (2004), Rovelli and Vidotto (2015; 2022), Vidotto (2017). While we agree in using the term 'extended configuration space' for referring to the configuration space of the parametrized version of the Newtonian particle, they use it also for the configuration space of non-deparametrizable models, which I will argue below is not extended in the sense that it does not include time variables or spacetime coordinates.

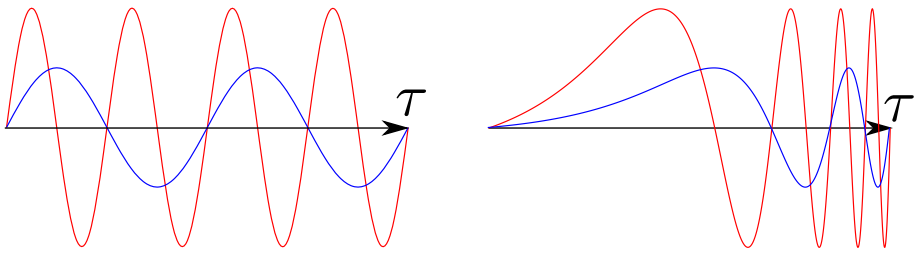

**Fig. 1** Representation of two equivalent solutions of the equations of motion of the double oscillator model. The two solutions define the same sequence of configurations but disagree on the value of the time coordinate $\tau$ they assign to each configuration

Non-deparametrizable, reparametrization invariant models, on the other hand, are reparametrization invariant models that are not defined on an extended configuration space. That is, all the variables in the configuration space of the model are variables with a physical meaning different from being a time parameter or a spacetime coordinate. The two paradigmatic examples of this class of models are general relativity and the Jacobi action for a Newtonian system.[28] Minisuperspace models also belong to this class, as the configuration space variables for these models are just the scale factor $a$, the scalar field $\phi$, and the lapse function $N$. Although $N$ is related to time, it is not a time variable, and neither is the scale factor nor the field.

Let me introduce an example of non-deparametrizable model to have a clearer picture of the properties of this class of models. This example is the Jacobi action for the double harmonic oscillator as discussed in Mozota Frauca (2023). In this model one describes the evolution of the position of two harmonic oscillators $x$, $y$ with respect to an arbitrary temporal parametrization $\tau$. Different solutions of the equations of motion of this model, for the same initial conditions, give the same trajectory in configuration space, although parametrized in different manners. This is illustrated in Fig. 1. The physical content of the model can be argued to be contained in the trajectory in configuration space and not in the way it is parametrized. For instance, the configuration space trajectory between two points contains information about how many times the oscillators oscillate or about the sequence of positions that the second oscillator takes every time the first one reaches its maximum amplitude. Note that during half an oscillation we can use one of the oscillators as a clock, but for the whole evolution it doesn't make sense to try to express the dynamics in relative terms as $x(y)$ or $y(x)$, as neither $x$ nor $y$ uniquely determines a moment of the evolution.

If we wanted to recover Newtonian time, there are ways of doing so, but for this class of systems, it is not a configuration (or phase) space function, so no deparametrization is possible. Instead, one needs to select a preferred way of parametrizing the configuration space trajectories, and for the case of the double harmonic oscillator one can show that the Newtonian parametrization allows defining a conserved quantity, energy, and recovering the Newtonian equations of motion. For the case of minisuperspace models, (proper) time is not an internal variable, but it can be recovered from a trajectory in minisuperspace by knowing the action of the system.

---

[28]The relevance of systems described by the Jacobi action for the problem of time was first noticed by Barbour (1994).

A crucial ingredient for the distinction between deparametrizable and non-deparametrizable models and for my argument in this paper is the concept of a time variable or spacetime coordinate. I take the temporal structure of a classical model to consist of two ingredients: an order relation and a metric aspect.[29] The order relation can be an order relation between different instants of the evolution of a system or a partial order relation between events for field theories.[30] Given the simplicity of the models I am studying in this article, I will leave the subtleties of field theories and full general relativity aside. For the simple classical models discussed here, one effectively has an absolute order relation between instants of the evolution, be it because we are dealing with Newtonian theories, or because we have a preferred foliation for the case of cosmological models. In this sense, the model specifies whether a pair $a, \phi$ comes before or after another $a', \phi'$ in the same trajectory. The metric aspect allows us to define the duration of temporal intervals, and it is captured by either the Newtonian time or the line element (Eq. 1). In this sense, I take that Newtonian time and proper time in the cosmological setting are good time variables, as they capture both the order and metric aspects of time.

Given this notion of time variables, it is important to distinguish them from a conceptual point of view from monotonic physical variables. A monotonic physical variable can be used for identifying instants of the evolution of a system, i.e., it can work as a clock, but it plays a different role from the temporal structure of the model. For instance, the position of a particle moving freely in Newtonian physics can be used for keeping track of time, but this is different from being time, which would still be there in a model in which the particle happened to be oscillating or didn't exist at all. I will return to this point in the next section.

Coming back to the distinction between deparametrizable and non-deparametrizable models, the goal of this section was to argue that both suffer from a problem of time, but that this problem can in principle be solved for the former class of models, while it seems unsolvable for the latter. Again, I refer the reader to Mozota Frauca (2023) for the technical details, but it can be shown that every reparametrization invariant model has a total Hamiltonian which is a combination of constraints, just as happened in the case of the minisuperspace model (Eq. 2). This means that when we apply the quantization procedure outlined in Sect. 2.2 we find a similar situation. That is, step 3 requires us to define physical states that satisfy the constraints, but, as the Hamiltonian is a combination of constraints, this implies that the action of the Hamiltonian on these states leaves them unchanged, and we are unable to obtain a non-trivial dynamical equation in step 4 of the quantization procedure. For the examples in this section, this means that instead of obtaining states $\psi(x, t, \tau)$ and $\psi(x, y, \tau)$, what we obtain is states of the form $\psi(x, t)$ and $\psi(x, y)$. There is a problem of time in both cases because the expected $\tau$-dependence is not there, although I will now argue that the problem is quite different in both cases.

---

[29] See a more extended version of this argument in Mozota Frauca (2024).

[30] In this paper I will discuss the order relation associated with time, but I do not want to make any claims about an arrow of time, about the past being different from the future, or anything along these lines. I take the order relation to be something compatible both with views that accept a temporal arrow and with views that deny it.

Consider first the example of the parametrized Newtonian particle. The canonical quantization of such a model defines states $\psi(x, t)$ satisfying a constraint equation that turns out to be the Schrödinger equation. In this case, it makes sense to use the internal time resolution of the problem of time discussed above and identify $t$ as an 'internal time', as it was time all along. Similarly, other resolutions of the problem of time for this model are based on the same identification.[31] This strategy works for any deparametrizable model, as, by definition, time is a variable in the (extended) configuration space of this kind of model and we can reinterpret wavefunctions satisfying the constraint equation in the extended configuration space as just wavefunctions that satisfy a dynamical equation which describe how they evolve with respect to $t$.

When we move to consider the example of the non-deparametrizable model, the double harmonic oscillator, what we find is bad news. The canonical quantization procedure gives us states $\psi(x, y)$ which are solutions of the time-independent Schrödinger equation. Here, it doesn't seem to make sense to appeal to the internal time strategy as neither $x$ nor $y$ seems to represent internal clocks or internal time variables. As discussed above, they are not even monotonic to start with. Forcing the internal time resolution would seem to identify one of the oscillators with time, and identify positions of the oscillator with instants of time. This seems wrong from a classical perspective, as there are many, infinitely many indeed, instants of the evolution in which an oscillator takes a given position (inside its range of oscillation). Moreover, time was an order relation between configurations of the whole system, and now the quantum theory seems to be taking one variable as physical and quantum and the other as a time variable, contrary to our expectations. For generic non-deparametrizable models, the same will hold: we will have a state satisfying certain constraint equations, but it would be wrong to interpret these states as wavefunctions describing how a state for some degrees of freedom evolves with respect to some internal time, as in this class of models no configuration space variable can be interpreted to be representing time.

Relevantly for the argument in this article, facts about duration are also missing in the canonical quantization of non-deparametrizable reparametrization invariant models. For classical models like the double harmonic oscillator model we could make claims like 'the system takes a time $t_0$ to go from configuration $x_1, y_1$ to configuration $x_2, y_2$'. From the simple formalism in which we just have wavefunctions $\psi(x, y)$ with no $t$ dependence it seems that this sort of claim is not recoverable. Given that we don't even have a well-motivated sense of evolution, it does not make sense to read durations from our model. And even if we consider the ill-motivated internal time strategy, facts about duration are at most recovered in an ad hoc and challengeable manner.

For instance, in the case of the double harmonic oscillator one could say that $\psi(x_1, y)$ and $\psi(x_2, y)$ represent the state for the second oscillator at the moments at which the first oscillator is at positions $x_1$ and $x_2$. Nothing in the formalism then tells us how much time elapses in between those two instants (which would be something an external system could be able to observe). One could, by fiat, impose that this time has to be the same as in the classical theory (for some initial velocity that would need to be somehow determined). This again goes against our intuitions, as one would have expected that the relationship between $x$ and $t$ wouldn't be fixed, but be subject to some quantum dynamics, such that it isn't fixed that $x$ goes from $x_1$ to $x_2$ in time $t_0$, but that there is some probability of it happening in this time, or maybe that it takes some other duration.

---

[31] See Mozota Frauca (2023) for a discussion of the different resolutions and their relation with deparametrization.

The other main interpretation of timeless states, the probabilistic interpretation that would read $\psi(x, y)$ as encoding some probability, is equally problematic, as we are not given a way to derive claims like 'the system takes a time $t_0$ to go from configuration $x_1, y_1$ to configuration $x_2, y_2$' from claims such as 'the probability for the oscillators to be at the configuration $x, y$ is $|\psi(x, y)|^2$' or 'the probability for the oscillators to go from $x_1, y_1$ to configuration $x_2, y_2$ is $p(x_1, y_1; x_2, y_2)$'.[32]

In this way, we see how the quantization of non-deparametrizable models lead to quantum models in which temporal structures disappear and that the main ways in which the formalism has been interpreted by the quantum gravity community fail to give an explanation or justification that would allow to recover, in the right conditions, the empirical content that in the timeful formalism we explained making use of temporal structures. To insist on the analogy in the introduction, this would be as if, when eliminating absolute simultaneity in going to a relativistic theory, we were unable to explain interactions between bodies. The structure of relativistic theories is sufficiently similar to the structure of Newtonian models so that the empirical content of the latter can be represented, up to small modifications, by making use of the former. Of course, from a Newtonian perspective one may have thought that absolute simultaneity is an empirical fact, but when revising our experimental evidence, together with the experiments that motivated relativity, we can convince ourselves that this is not the case. Moreover, relativity is also able to explain why, even in a world with no absolute simultaneity, in some situations it may look like there is.

In the case of the problem of time we are in a situation in which one of the two theoretical structures is clearly less rich, and the representation of the old theory in terms of the new one seems impossible. The arguments above support the claim that by making use of $\psi(x, y)$ one cannot recover $x(t), y(t)$. At least the interpretations proposed by the community fail, as argued in more detail and generality in Mozota Frauca (2023). Now, the analogy with the absolute simultaneity case may make the idea of dropping temporal claims as predictions of the model tempting, but this position is unattractive, both for deparametrizable models in general and for cosmological models in particular. I will now argue for this claim in more detail.

## 4 Is the Age of the Universe (or Duration in General) a Physical Prediction?

In the previous section I have built on the argument in Mozota Frauca (2023; 2024) to argue that the canonical quantization of minisuperspace models in quantum cosmology fails to lead to a satisfactory quantum theory and that a way of arguing for this is that the states $\psi(a, \phi)$ do not have information about $t$, which had a physical and empirical meaning in the classical theory. In this section I will further argue that $t$ has physical meaning and that its disappearance in the quantum theory is problematic. The argument in this section is to be contrasted with the discussions by some authors in the quantum gravity community, who tend to dismiss (even if sometimes implicitly) the physical content encoded in $t$. In this way, the attitude towards the problem of time in quantum cosmology by this community is that if $t$ goes missing, that's not too serious, as they take the content of the models (even classical ones) to be exhausted by $a$ and $\phi$.

---

[32] One would need to specify the way in which the probability $p(x_1, y_1; x_2, y_2)$ would have to be computed. Rovelli (2004) and Colosi and Rovelli (2003) propose to use the inner product structure of the physical Hilbert space for this.

In the first place, cosmology's being a relativistic theory, one may worry that my insistence on $t$ could be contrary to the spirit of relativity, as it could seem that I am picking a preferred time variable. But I want to insist that this is not the case. I claim that $t$ has a physical meaning: it is the proper time that a class of observers, those at rest with respect to the isotropic and homogeneous spaces in which the FLRW can be decomposed, could measure. This physical meaning is of course independent of any choice of spacetime coordinates, so even if we choose a coordinate system such that $t$ isn't one of our coordinates, it still has the same physical meaning. The choice of class of observers could also be questioned on the grounds of relativity, but, again, this choice is just a matter of convenience and simplicity. The proper time of any observer is physically meaningful, not just that of observers at rest. In this sense, the argument in this paper works even if we choose any other observer. Imagine in the classical theory an observer in a spacecraft traveling along a very chaotic, non-geodesic trajectory in the FLRW spacetime. The proper time this space-traveler experiences, as can be computed from Eq. (1) is also part of the physical content of the classical model. This physical prediction of the classical model is also missing from the quantum state $\psi(a, \phi)$, just as $t$ was missing. In this sense, let me insist that my claim that $t$-related facts are predictions of the classical model that are lost in the quantum one is compatible with relativity, as there are also $t'$-related facts that are lost, for the proper time $t'$ of any observer.

The other main sort of objection would come from some relational views about time that are common in the quantum gravity literature. These views are based on observations like the fact that we never directly measure time, but we use physical clocks, made of material systems for keeping track of time. Given this fact, one can put some pressure on the physicality of time variables, and, more specifically, of $t$ in the cosmological models we are concerned with in this paper. I will distinguish two sorts of relationalism about time that play a role in debates in quantum gravity, I will analyze what they claim for our models, and I will nevertheless conclude that $t$ has a physical meaning that is not captured by the quantum models and that relationalists should account for.

First, there are Machian or neo-Machian versions of temporal relationalism as have been put forward by Julian Barbour and collaborators.[33] Roughly speaking, this version of relationalism denies that there is a preferred way of assigning time labels to a trajectory in the configuration space of the universe. This criticism directly applies to Newtonian mechanics and motivates shape dynamics, a close relative to general relativity in which there is no absolute time, even if there is something like a privileged foliation. For the case of Newtonian mechanics, we could summarize the Machian view as saying that if we had a model describing all the particles of the universe it would be enough to have a trajectory in configuration space for having all the physical content of the theory. That is, this view denies the physical meaning of the metric aspect of Newtonian time, as the trajectory in configuration space already contains the information of the 'time' each material clock would be indicat-ing at each moment in the trajectory. The argument for general relativity or shape dynamics is more sophisticated, but it similarly denies the metric aspect of time in general relativity.

---

[33] See for instance Barbour (1994; 2011).

Does this view about time affect my analysis that $t$ carries physical information? I will now argue it doesn't. In the Newtonian example, one could eliminate absolute time by realizing that the information about every clock was represented by the trajectory of the configuration space, as it described a sequence of configurations for every degree of freedom in the universe. Minisuperspace models, however, do not necessarily share this pretension of including every degree of freedom in the universe, and one can see $t$ as encoding how some physical systems behave or would behave, even if approximately. For instance, a minisuperspace model can be used for modeling our universe, and even if a great deal of physical details and fine structure is left out, the variable $t$ is useful for describing the behavior of systems like atomic clocks far from galaxies or densely populated regions of spacetime. That is, we do not need to explicitly add to a minisuperspace model the details about atomic clocks and their dynamics to be able to describe how they will evolve. It is $t$, or more generally, the line element (Eq. 1), which determines how physical clocks will behave without needing to introduce them explicitly in our models. In this sense, one could agree with the general spirit of the Machian argument in that if a complete description of all the degrees of freedom of the universe were in place, then the metric aspect of time wouldn't be necessary but still hold that $t$, or the line element (Eq. 1), has some physical meaning, as minisuperspace models do not generally aim to capture every aspect of a universe.

Notice also that the Machian argument affects the metric aspect of time but not its order aspect, so it supports a relational view of time but in a way that doesn't completely eliminate temporal structure. In this sense, even if the Machians could deny the importance of the metric aspect of time, they could still find it worrisome that the order structure disappears or is radically affected when quantizing a theory. This is precisely what happens in the quantization of non-deparametrizable reparametrization invariant models. The Jacobi action for the double harmonic oscillator as described in Mozota Frauca (2023) and discussed in the previous section is a good example of the Machian spirit.[34] The temporal structure of the model doesn't define a preferred metric for time, but the order relation is perfectly defined. However, when quantizing the theory this order just disappears, and, as I argued in the previous section, it seems wrong to replace such an order with an order defined in a forced way by taking one of the oscillators to be something like an internal time. The same applies to minisuperspace models, where we have a clear distinction in the classical theory between the physical degrees of freedom $a, \phi$ and the temporal structure of the model. In this sense, one can be Machian about duration or absolute time scales but still find the lack of temporal structure of canonical quantum cosmology problematic.

To summarize my view about how Machian critiques affect the argument in this article, let me insist on two claims. First, the metric aspect of $t$ in a minisuperspace model captures something that the Machian would want to preserve, i.e., even the Machian would agree that there is something physically and empirically meaningful in claiming that the universe is 13.8 billion years old (even if this claim needs to be translated and put in relation to how systems like atomic clocks behave). Second, the disappearance of the order structure of the classical model in the quantum one should be equally worrying for the Machians, as they agree that temporal order has a physical meaning.

---

[34] Indeed, the Jacobi action of Newtonian systems is widely discussed in this literature (Barbour 1994; Gryb 2010).

Let me now move to another sort of relationalism which is more popular in the quantum gravity literature and which is more radical.[35] This brand of relationalism claims that, even in general relativity, the physical content of a theory lies in the correlations between 'observables' it defines.[36] This view seems to deny, or at least it rarely if ever mentions it, the order aspect of temporality. In this sense, I will take this sort of relationalism, or positions inspired by it, to be denying both the order and the metric aspect of time as represented by $t$ in minisuperspace models.[37] If there is a different reading of this type of relationalism that doesn't deny that there are physical facts captured by the order and metric temporal structures of minisuperspace models, then it does not represent an objection to my argument.

For minisuperspace models like the one discussed in this article, there would be only two 'observables' according to this relationalism, $a$ and $\phi$, and the content of these models would lie in the relations between them, that is, in $a(\phi)$ or $\phi(a)$. This is made explicit by Rovelli (2004, 298)[38]:

> The physical content of the theory [a minisuperspace model] is not in the dependence of these two quantities [$a$ and $\phi$] on $t$, but in their dependence on each other. The proper meaning of (8.4)–(8.9) [the equations for $a(t)$ and $\phi(t)$] concerns the relation between $\phi$ and $a$.

This interpretation of the classical model is in my opinion problematic, as it leaves out the physical content encoded in $t$. Even at a classical level, relationalism as defended by Rovelli cannot read the age of the universe from the variable $t$.

Let me consider what this sort of relationalism would say about an example like the double harmonic oscillator discussed in the previous section. If it is just correlations between the observables that matter, the physical content of the model would be given by just the set of pairs $x$, $y$ that constitute a physically allowed trajectory in phase space, but without specifying any information about which pair comes after or before another. In this sense, we would be disregarding putative predictions of the model like what the position of the second oscillator would be the next time the first one is at its maximal position or after 5 oscillations of that oscillator. That is, as the correlations give for each value $x_0$ a set of allowed values of $y$ and not an ordered set of them, the relationalist could consider any information about which comes before or after as unphysical. Is this interpretation sound? In my opinion, it depends on what one is taking the model to represent, but the most natural interpretation, at least when we move to consider the cosmological model at least, is that the order relation that the model defines is physically relevant.

---

[35] This sort of relationalism has been most famously defended by Carlo Rovelli and Francesca Vidotto: Rovelli (2004), Rovelli and Vidotto (2015; 2022), Vidotto (2017).

[36] It is beyond the scope of this paper to analyze the motivation for this view, which is partly inspired by a particular analysis of the reparametrization invariance of general relativity. I refer the reader to Maudlin (2002), Pitts (2014; 2017), and Mozota Frauca (2023; 2024) for views critical with this analysis of general relativity and this sort of relationalism.

[37] This reading of relationalism as presented by Rovelli can be found in Thebault (2012) when it is claimed that "the RDT [Rovelli-Dittrich-Thiemann] approach can only naturally be interpreted in terms of a philosophical framework which precludes temporal structure altogether" (Thebault 2012, 290). See also the discussion in Thebault (2021) and Mozota Frauca (2025).

[38] See also the discussion of minisuperspace models in Vidotto (2017).

A way in which one can interpret the double oscillator model such that order predictions are unphysical is the following. Imagine that we have a random machine that, for some reason, gives as an output pairs $x$, $y$ which always correspond to points in a given trajectory of our model of the double harmonic oscillator. Each output is completely independent of the one before, and for predicting the next output one only knows that it will be part of the same trajectory. In this case, the order relation of the model is quite useless, and one could regard it as unphysical. Conversely, one can think of the model as representing a set of two physical harmonic oscillators embedded in spacetime that we can go and observe. By assumption, there will be physical facts about the sequence of configurations of the system that an observer would observe, and these physical facts are captured by the order relation in the model. The interpretation I think is more sensible of a classical minisuperspace model is more similar to this latter interpretation than to the machine that spits out random pairs $x$, $y$.

An objection that the relationalist could raise against this argument is that by embedding the double harmonic oscillator system in a spacetime and with the presence of observers one is 'cheating', in the sense that one is enlarging the physical system and that it is really correlations $x$, $y$, $w$ that are physically meaningful, where $w$ represents the state of the environment, the observer or any other physical system we may want to add. In this sense, the relationalist could try denying the physical relevance of order relations of the bigger system and interpret the order relations of the smaller system just as encoding correlations with an external system. One could go on and discuss the meaningfulness of order relations when the system in question is the whole universe and they cannot be claimed to represent correlations with something external, but this is not necessary for the argument in this article. Indeed, above I have argued that minisuperspace models can be interpreted not as aiming to give a complete description of the universe, but just an approximate one that concerns a few of its degrees of freedom.[39] Accepting this view, it is reasonable to accept that $t$ and the order relation of minisuperspace models capture facts about what observers in this universe would observe or about the evolution of physical systems like atomic clocks. In this sense, I believe relationalists would have to accept that the order relation of minisuperspace models has a physical meaning, even if they would rather express it in terms of correlations between the degrees of freedom contained in the model and not in terms of an order relation.

Similarly, I believe that relationalists would have to accept that there is some physical meaning in the metric aspect of $t$. $t$ can be seen as encoding the time that an atomic clock would be indicating, even if approximately. Some relationalists could object that there is no such a clock, and would probably be right, but we have plenty of astronomical objects in our universe that we can use to keep track of $t$. For instance, we have solid theories describing stars and how they evolve, and we can take that knowledge to use stars as clocks. Take for instance the claim that the universe is not old enough to have formed supermassive black holes via stellar collapse[40] This claim is a physically meaningful claim that the relationalists would better incorporate into their account of the temporal structure of the universe. Here I am not arguing that it is not possible for the relationalist to do so, but just that $t$, in a minisuperspace model captures this.[41] In this sense, I believe that for the relationalist story to be able to account for all the physical content of models like the cosmological models we are considering here, it'd better include, in some way or another, the physical content encoded by $t$.

---

[39] This attitude is also held explicitly in Vidotto (2017) by one of the proponents of relationalism.

[40] This and similar claims can be found in the cosmology and astrophysics literature. See for instance Volonteri (2012).

Coming back to the reading that Rovelli makes of minisuperspace models, I believe that my argument shows that the relations between $a$ and $\phi$ do not exhaust all the physical content of these models. These models not only predict that a universe of size $a_1$ with a matter content $\phi_1$ will expand to a size $a_2$ with a matter content $\phi_2$, but also how quickly it will do so. If relationalists like Rovelli accept this, then they will be forced to include $t$ among what they consider 'observable'.

For these reasons, I believe that relationalism, Machian or otherwise, cannot deny the physical meaning that $t$ has in our classical cosmological models, and in minisuperspace models in particular. $t$, in both its order and metric aspects, is a key ingredient of the best accounts we have for explaining a wide variety of cosmological and astrophysical phenomena. Perhaps there is a relationalist account that can dispense of $t$ once all the degrees of freedom of the universe are considered, but for minisuperspace models it is not the case that we can dispense of $t$ without missing important physical information of our classical cosmological model.

If the relationalists agree with me that there is something physical related to $t$ in the classical model, then they would have to agree that it is prima facie worrisome that it altogether disappears in the canonical quantization of the model. The canonical quantization of the model does not contain information about $t$, and hence it seems unable to make predictions about the duration of different intervals of the universe. If the order structure was defined by $t$ and the dynamical variables were $a$ and $\phi$, taking $a$ or $\phi$ to be defining the temporal structure and the other variable to be dynamical as is done in the internal time interpretation of canonical quantum cosmological models should be question-begging. Even for the most radical of relationalists, there must be something suspicious in going from the triple $a, \phi, t$ to just the couple $a, \phi$.

Again, let me insist that the worry is not that we are expecting to find exactly the same structures in the quantum version of the theory. The history of physics teaches us that our picture of the world changes with theory change and one should keep an open mind. The worry is that the new theory should be able to represent, even if in its own terms, the empirical content of the old one. As I have just argued, we have good reasons for believing that $t$ in minisuperspace models has empirical meaning, and relationalist intuitions are not strong enough for us to dismiss it. Therefore, the fact that time is missing from the formalism becomes problematic if our interpretations of this formalism fail to recover the empirical content associated with it. Let me insist in the ways in which the main interpretations of minisuperspace models fail to do so.

First, a way a defender of the internal time interpretation could try to recover predictions like the age of the universe is by postulating that the internal time is related in some way to the time $t$ of the classical model. As in the classical model we are considering $a(t)$ and $\phi(t)$ to be monotonic functions of $t$, one could invert this relationship and take $a$ or $\phi$ to be determining $t$. In this sense, if in the internal time interpretation we could take $\psi(a, \phi)$ to be a wavefunction for $a$ evolving with respect to $\phi$, by making use of the classical relationship between $\phi$ and $t$ we could convert it into a wavefunction of the form $\psi(a, t)$. However, this strategy does not seem very promising. In the first place, it is based on the contingent fact that in the model we are considering $a$ and $\phi$ to be monotonic, but we could change the model so that this is not the case, and indeed we study a realistic model of the universe is very likely that there isn't any such variable.[42] Moreover, even if we were granted that there is a monotonic variable, its relation with $t$ would also be contingent. Different matter contents imply different expansion rates and the detail of the evolution of matter fields also depends on non-fixed facts like initial conditions. Therefore, importing to the quantum theory the relationship between the internal clock and real time seems question-begging. Second, one can insist in the general objection against the internal time strategy that both $a$ and $\phi$ are dynamical variables in the classical theory, and fixing one of them to represent temporal structure in the quantum model would mean that there are no quantum phenomena associated with it.

---

[41] Let me emphasize how powerful minisuperspace models are. While for the relationalist one needs to explicitly introduce external systems in the model in order to make predictions about them (if they want to dismiss the meaning of $t$), in the standard reading of the model these predictions are straightforward.

Second, let me comment on probabilistic interpretations of the canonical quantization of minisuperspace models. In these interpretations one takes $\psi(a, \phi)$ to be encoding something like the probability of measuring some values $a$ and $\phi$, or one builds on the inner product structure of the Hilbert space to produce objects of the form $p(a, \phi; a', \phi')$ which are interpreted to encode the probability of finding $a, \phi$ given that $a', \phi'$ were observed before.[43] It is unclear to me how to make sense of these claims and how to give meaning to such probabilities. By claiming that these structures define probabilities, the picture of reality that is given seems closer to that of the machine spitting random pairs $a, \phi$ and not to the classical picture of cosmology, in which there were well-defined temporal structures. In my opinion, there is a gap that needs to be bridged between both pictures. For instance, it seems to me that if one were able to make the quantum 'prediction' that 'the probability of measuring a scale factor $a_1$ and a scalar field $\phi_1$ after having observed $a_0, \phi_0$ is of 98 %' there will be something that has got lost from the classical claim that 'the universe goes from $a_0, \phi_0$ to $a_1, \phi_1$ in 13.8 billion years'. In the same way that claims about $t$ seem to be lost in the internal time interpretation of quantum cosmology, they seem to be equally lost in probabilistic resolutions.

Therefore, we see how the two main interpretations of quantum cosmological models fail to recover the empirical content represented by $t$ in the classical model. Relationalists who accept that $t$ has some physical meaning would have to accept that there is some problem with quantum minisuperspace models. Notice that for a relationalist, if they accept my arguments, $t$, or temporal structure in general, has a physical meaning in any model used to describe just some of the degrees of freedom of the universe and not all of it. In this sense, they may hope that the problem of time dissolves once the model one quantizes represents every single degree of freedom in the universe. I believe that there are some conceptual troubles with this position, but it is beyond the scope of this article to discuss them.

---

[42] In our expanding cosmology, the scale factor would be the best candidate, but it is at least conceivable, and we have models that predict that the universe underwent or will go through contracting phases.

[43] The former type of interpretation includes the 'naive Schrödinger' and the 'conditional probability' interpretations as described in the review Kuchar (1992), while the latter is the more recent transition amplitude interpretation as defended in Rovelli (2004), Rovelli and Vidotto (2015; 2022), and Vidotto (2017).

## 5 Conclusion

In this article I have argued that the quantum cosmological models derived by applying the canonical quantization procedure to classical minisuperspace models are affected by the problem of time in a way that can make us suspect that there is something wrong with this quantization procedure for such models. Indeed, I have argued that while classical cosmology allowed us to make claims like 'the age of the universe is 13.8 billion years' or 'the universe is not old enough for having produced supermassive black holes via stellar collapse', this sort of claims is missing from canonical quantum cosmology. I have argued that this is not like in some other cases of theory change in which empirical claims were kept, even if adapted to a new formalism and conceptual schema. In the case of quantum cosmological models, I have argued that the main interpretations of the formalism fail to preserve this empirical content, not even in some appropriately adapted way.

I have paid attention to popular views in the quantum gravity community, which use relationalist arguments to dismiss the physical and empirical meaning of temporal structures even in our classical theories. Contrary to these views, I have argued that temporal claims in our classical models constitute empirical content that a quantum model should be able to explain, even if in its own terms.

In particular, I have argued that internal clock resolutions of the problem of time and probabilistic interpretations of the formalism are equally unsatisfactory and unable to recover the sort of information that in the classical models was encoded in $t$. I haven't discussed in much detail the semiclassical approaches to cosmology, but I have noted that even if one can postulate temporal structures in these approaches,[44] the way this is done is simply not applicable for generic states and we are therefore missing a sound and consistent way of interpreting canonical quantum cosmological models which overcomes the conceptual issues associated with the problem of time.

The discussion in this article connects with the general discussion of the problem of time of canonical approaches to quantum gravity. If one accepts the conclusion of this article, then one should at least contemplate the possibility that the problem of time also makes canonical approaches to quantum gravity unsatisfactory. In this sense, my conclusion in this article agrees with the conclusion reached in Mozota Frauca (2023), namely that for non-deparametrizable reparametrization invariant models there is a problem of time which cannot be solved in a similar fashion to the way it was solved in deparametrizable models. Mozota Frauca (2023) argued that this affected general relativity, and this article extends the conclusion to cosmological models.

An analysis in depth of the possible resolutions of this problem is beyond the scope of this article, but let me just mention a couple of ideas. First, one could try to aim for a deparametrizable extension or version of general relativity, which is the idea pursued by the unimodular gravity program.[45] Second, one could look for a different quantization method, as the relational quantization proposal in Gryb and Thebault (2012; 2016) and applied to a cosmological setting in Gryb and Thebault (2019). Both ideas have to face challenges and it would be interesting to study to which extent the models they lead to are able to recover the age of the universe.

---

[44]The way in which this is done and the justification for it may in my opinion be challenged.

[45]Again, see Brown and York (1989) for the paper which inspired the approach and Kuchar (1992) and Isham (1993) for some criticisms of it.

**Acknowledgements**  I want to thank the audiences at Berlin, Helsinki and Milan for their useful feedback.

**Author Contributions**  All the authors contributed equally to this work.

**Funding**  Open Access funding provided thanks to the CRUE-CSIC agreement with Springer Nature.

# References

Ade, P.A.R., et al. 2016. Planck 2015 results - XIII. Cosmological parameters. *Astronomy & Astrophysics* 594 : 13. https://doi.org/10.1051/0004-6361/201525830. https://arxiv.org/abs/1502.01589

Aghanim, N., et al. 2020. Planck 2018 results - VI. Cosmological parameters. *Astronomy & Astrophysics* 641 : 6. https://doi.org/10.1051/0004-6361/201833910. https://arxiv.org/abs/1807.06209.

Agullo, I., and P. Singh. 2017. Loop quantum cosmology. In *Loop quantum gravity: the first 30 years*, 183–240. Singapore: World Scientific. https://doi.org/10.1142/9789813220003_0007.

Anderson, E. 2017. *The problem of time*, vol. 190. Cham: Springer. https://doi.org/10.1007/978-3-319-58848-3.

Ashtekar, A. 2009. Loop quantum cosmology: An overview. *General Relativity and Gravitation* 41 (4): 707–741. https://doi.org/10.1007/s10714-009-0763-4.

Ashtekar, A., and P. Singh. 2011. Loop quantum cosmology: A status report. *Classical and Quantum Gravity*. https://doi.org/10.1088/0264-9381/28/21/213001.

Ashtekar, A., T. Pawlowski, and P. Singh. 2006. Quantum nature of the big bang: An analytical and numerical investigation. *Physical Review D Particles, Fields, Gravitation and Cosmology* 73 (12): 124038. https://doi.org/10.1103/PhysRevD.73.124038.

Ashtekar, A., Pawlowski, T., and Singh, P. 2006. Quantum nature of the big bang: Improved dynamics. *Physical Review D Particles, Fields, Gravitation and Cosmology* 74 (8): 084003. https://doi.org/10.1103/PhysRevD.74.084003.

Barbour, J. B. 1994. The timelessness of quantum gravity: I. The evidence from the classical theory. *Classical and Quantum Gravity* 11 (12): 2853. https://doi.org/10.1088/0264-9381/11/12/005.

Barbour, J. B. 2011. Shape Dynamics. An Introduction. In *Quantum Field Theory and Gravity*, 257–297. New York: Springer. https://doi.org/10.48550/arxiv.1105.0183.

Bojowald, M. 2011. Quantum cosmology: A fundamental description of the universe. *Lecture Notes in Physics* 835: 1–294. https://doi.org/10.1007/978-1-4419-8276-6.

Bojowald, M. 2015. Quantum cosmology: A review. *Reports on Progress in Physics* 78 (2): 023901. https://doi.org/10.1088/0034-4885/78/2/023901.

Bojowald, M., and H. A. Morales-Técotl. 2004. Cosmological applications of loop quantum gravity. In *The early universe and observational cosmology*, 421–462. New York: Springer.

Brown, J. D., and J. W. York. 1989. Jacobi's action and the recovery of time in general relativity. *Physical Review D* 40 (10): 3312. https://doi.org/10.1103/PhysRevD.40.3312.

Calcagni, G. 2017. *Classical and quantum cosmology*. New York: Springer. https://doi.org/10.1007/978-3-319-41127-9.

Callender, C., and C. D. McCoy. 2021. Time in cosmology. In *The Routledge companion to philosophy of physics*, 707–718. Milton Park: Routledge. https://doi.org/10.4324/9781315623818-66.

Chua, E. Y. S., and C. Callender. 2021. No time for time from no-time. *Philosophy of Science* 88 (5): 1172–1184. https://doi.org/10.1086/714870.

Colosi, D., and C. Rovelli. 2003. Simple background-independent Hamiltonian quantum model. *Physical Review D Particles, Fields, Gravitation and Cosmology*. https://doi.org/10.1103/PhysRevD.68.104008.

Dirac, P. A. M. 1964. *Lectures on quantum mechanics*. New York: Belfer Graduate School of Science Yeshiva University.

Gerlach, U. H. 1969. Derivation of the ten Einstein field equations from the semiclassical approximation to quantum geometrodynamics. *Physical Review* 177 (5): 1929. https://doi.org/10.1103/PhysRev.177.1929.

Gielen, S., and L. Menéndez-Pidal. 2022. Unitarity, clock dependence and quantum recollapse in quantum cosmology. *Classical and quantum gravity* 39 (7): 075011. https://doi.org/10.1088/1361-6382/AC504F.

Gielen, S., and L. Menéndez-Pidal. 2022. Unitarity and quantum resolution of gravitational singularities. *International Journal of Modern Physics D*. https://doi.org/10.1142/S021827182241005X.

Gryb, S. 2010. Jacobi's principle and the disappearance of time. *Physical Review D* 81(4): 044035. https://doi.org/10.1103/PhysRevD.81.044035.

Gryb, S., and K. P. Y. Thebault. 2012. The role of time in relational quantum theories. *Foundations of Physics* 42 (9): 1210–1238. https://doi.org/10.1007/S10701-012-9665-5/FIGURES/2.

Gryb, S., and K. P. Y. Thebault. 2016. Time remains. *The British Journal for the Philosophy of Science* 67 (3): 663–705. https://doi.org/10.1093/bjps/axv009.

Gryb, S., and K. P. Y. Thebault. 2019. Bouncing unitary cosmology I. Mini-superspace general solution. *Classical and Quantum Gravity* 36 (3): 035009. https://doi.org/10.1088/1361-6382/AAF823.

Halliwell, J. J. 1990. Introductory lectures on quantum cosmology. In *Quantum cosmology and baby universes*, ed. S. Coleman, J. B. Hartle, T. Piran, and S. Weinberg. Singapore: World Scientific.

Huggett, N., and Thebault, K. 2023. Finding time for Wheeler-Dewitt cosmology. https://arxiv.org/abs/2310.11072. https://philarchive.org/rec/HUGFTF.

Isham, C. J. 1993. Canonical quantum gravity and the problem of time. In Kunstatter, G., Vincent, D., Williams, J. *Integrable systems, quantum groups, and quantum field theories*, ed. L. A. Ibort, and M. A. Rodríguez, 157–287. https://doi.org/10.1007/978-94-011-1980-1_6. arXiv: gr-qc/9210011.

Kiefer, C. 2012. *Quantum gravity*. New York: Oxford University Press. https://doi.org/10.1093/acprof:oso/9780199585205.001.0001.

Kuchar, K. V. 1992. Time and interpretations of quantum gravity. In Kunstatter, G., Vincent, D., Williams, J. *Proceedings of the 4th Canadian Conference on General Relativity and Relativistic Astrophysics*, ed. G. Kunstatter, D. Vincent, and J. Williams. Singapore: World Scientific. https://doi.org/10.1142/S0218271811019347.

Maudlin, T. W. E. 2002. Thoroughly muddled McTaggart: Or, how to abuse gauge freedom to create metaphysical monstrosities. *Philosophers' Imprint* 2 (4): 1–23.

Mozota Frauca, A. 2023. Reassessing the problem of time of quantum gravity. *General Relativity and Gravitation* 55 (1): 21. https://doi.org/10.1007/s10714-023-03067-x.

Mozota Frauca, A. 2024. The problem of time for non-deparametrizable models and quantum gravity. In *Current topics in logic and the philosophy of science*, ed. F. Bianchini, V. Fano, P. Graziani. Papers from SILFS 2022 postgraduate conference. The SILFS series, vol. 4. College Publications. https://philsci-archive.pitt.edu/24052/.

Mozota Frauca, A. 2024. Time is order. In *Time and timelessness in fundamental physics and cosmology: historical, philosophical, and mathematical perspectives*, ed. S. De Bianchi, M. Forgione, and L. Marongiu, 49–67. Cham: Springer. https://doi.org/10.1007/978-3-031-61860-4_4.

Mozota Frauca, A. 2024. GPS observables in Newtonian spacetime or why we do not need physical coordinate systems. *European Journal for Philosophy of Science* 14 (4): 51. https://doi.org/10.1007/s13194-024-00611-7.

Mozota Frauca, A. 2025. Against radical relationalism: In defense of the ordinal structure of time. *Foundations of Physics* 55 (3): 37. https://doi.org/10.1007/s10701-025-00850-5.

Pitts, J. B. 2014. Change in Hamiltonian general relativity from the lack of a time-like killing vector field. *Studies in History and Philosophy of Science Part B: Studies in History and Philosophy of Modern Physics* 47: 68–89. https://doi.org/10.1016/j.shpsb.2014.05.007.

Pitts, J. B. 2017. Equivalent theories redefine Hamiltonian observables to exhibit change in general relativity. *Classical and Quantum Gravity* 34 (5): 1–23. https://doi.org/10.1088/1361-6382/aa5ce8.

Rothe, H. J., and K. D. Rothe. 2010. *Classical and quantum dynamics of constrained Hamiltonian systems*, vol. 81. Singapore: World Scientific. https://doi.org/10.1142/7689.

Rovelli, C. 2004. *Quantum gravity*. Cambridge: Cambridge University Press. https://doi.org/10.1017/CBO9780511755804.

Rovelli, C., and F. Vidotto. 2015. *Covariant Loop Quantum Gravity: An elementary introduction to quantum gravity and spinfoam theory*. Cambridge: Cambridge University Press. https://doi.org/10.1017/CBO9781107706910.

Rovelli, C., and Vidotto, F. 2022. Philosophical foundations of loop quantum gravity. https://arxiv.org/abs/2211.06718v2.

Smeenk, C. 2013. Time in cosmology. In *A companion to the philosophy of time*, 201–219. Hoboken: Wiley. https://doi.org/10.1002/9781118522097.CH13.

Thebault, K. P. Y. 2012. Three denials of time in the interpretation of canonical gravity. *Studies in History and Philosophy of Science Part B: Studies in History and Philosophy of Modern Physics* 43 (4): 277–294. https://doi.org/10.1016/J.SHPSB.2012.09.001.

Thebault, K .P. Y. 2021. The problem of time. In *The Routledge companion to philosophy of physics*, ed. E. Knox, and A. Wilson. New York: Routledge.

Unruh, W. G., and R. M. Wald. 1989. Time and the interpretation of canonical quantum gravity. *Physical Review D* 40 (8): 2598. https://doi.org/10.1103/PhysRevD.40.2598.

Vidotto, F. 2017. Relational Quantum Cosmology. In *The philosophy of cosmology*, 297–316. Cambridge: Cambridge University Press. https://doi.org/10.1017/9781316535783.016.

Volonteri, M. 2012. The formation and evolution of massive black holes. *Science* 337 (6094): 544–547. https://doi.org/10.1126/SCIENCE.1220843/SUPPL_FILE/1220843S2.MPG.

Weinberg, S. 2008. *Cosmology*. London: Oxford University Press.